\documentclass[draft,tightenlines,nofootinbib,preprint,aps,eqsecnum,amsmath,amssymb]{revtex4}

\newcommand{\beq}{\begin{equation}}
\newcommand{\eeq}{\end{equation}}
\newcommand{\bea}{\begin{eqnarray}}
\newcommand{\eea}{\end{eqnarray}}

\newcommand{\sgn}{\epsilon}

\begin{document}

\title{Dark Matter as a Relativistic Inertial Effect in Einstein Canonical
Gravity?}

\medskip

\author{Luca Lusanna}

\affiliation{ Sezione INFN di Firenze\\ Polo Scientifico\\ Via Sansone 1\\
50019 Sesto Fiorentino (FI), Italy\\ Phone: 0039-055-4572334\\
FAX: 0039-055-4572364\\ E-mail: lusanna@fi.infn.it}

\begin{abstract}

After the study of non-inertial frames in special relativity with
emphasis on the problem of clock synchronization (i.e. of how to
define 3-space), an overview is given of Einstein canonical gravity
in the York canonical basis and of its Hamiltonian Post-Minkowskian
(PM) linearization in 3-orthogonal gauges. It is shown that the York
time (the trace of the extrinsic curvature of 3-spaces) is the
inertial gauge variable describing the general relativistic remnant
of the clock synchronization gauge freedom. The dark matter implied
by the rotation curves of galaxies can be explained with a choice of
the York time implying a PM extension of the Newtonian celestial
frame ICRS.
\bigskip

Talk at "Group 28: The XXVIII International Colloquium on
Group-Theoretical Methods in Physics", Northumbria University,
Newcastle, 26-30 July 2010.

\today

\end{abstract}

\maketitle

\vfill\eject

We can speak of predictability in physics when there is a well-posed
Cauchy problem for the partial differential equations involved in
the problem under investigation. A pre-requisite for the Cauchy
problem is a sound definition of an instantaneous 3-space where the
Cauchy data are given. While in the Galilei space-time of Newtonian
physics this is not a problem due to the absolute nature of time and
of Euclidean 3-space, no intrinsic notion of 3-space exists in
special and general relativistic theories. In special relativity
(SR), where there is the absolute Minkowski space-time, only the
conformal structure (the light-cone) is intrinsically given. The
standard way out from the problem of 3-space is to choose the
Euclidean 3-space of an inertial frame centered on an inertial
observer and then use the kinematical Poincare' group to connect
different inertial frames. However, this is not possible in general
relativity (GR), where there is no absolute notion since also
space-time becomes dynamical (with the metric structure satisfying
Einstein's equations). The equivalence principle implies the absence
of global inertial frames: in the restricted class of globally
hyperbolic, asymptotically Minkowskian at spatial infinity,
space-times the best we can have are global non-inertial frames
connected by 4-diffeomorphisms (the gauge group of GR).
\bigskip

As a consequence, also in SR we have to face the problem of
reformulating physics in non-inertial frames centered on accelerated
observers \cite{1} as a first step before facing GR.\bigskip

In Minkowski space-time the Euclidean 3-spaces of the inertial
frames centered on an inertial observer A are identified by means of
Einstein convention for the synchronization of clocks: the inertial
observer A sends a ray of light at $x^o_i$ towards the (in general
accelerated) observer B; the ray is reflected towards A at a point P
of B world-line and then reabsorbed by A at $x^o_f$; by convention P
is synchronous with the mid-point between emission and absorption on
A's world-line, i.e. $x^o_P = x^o_i + {1\over 2}\, (x^o_f - x^o_i) =
{1\over 2}\, (x^o_i + x^o_f)$. Therefore the description of
non-inertial frames must replace Einstein convention with a more
general one allowing non-Euclidean 3-spaces.\bigskip

Since in GR the gauge freedom is the arbitrariness in the choice of
the 4-coordinates, in the non-inertial frames of SR a similar
arbitrariness is expected. However, the experimental description of
matter in both theories is based on {\it metrological conventions}:
a) an atomic clock as a standard of time; b) the velocity of light
in place of a standard of length; c) a conventional reference frame
centered on a given observer as a standard of space-time (think to
GPS!). The description of satellites around the Earth is done by
means of NASA coordinates \cite{2} either in ITRS2003 \cite{3}
(frame fixed on the Earth surface) or in GCRS (the geocentric frame,
centered on the Earth center, of IAU2000 \cite{4}). The description
of planets and spacecrafts uses the BCRS of IAU2000 \cite{4}
centered on the barycenter of the Solar System (a quasi-inertial
frame in a nearly Minkowski space-time due to gravity) Both GCRS and
BCRS are connected to Post-Newtonian solutions of Einstein equations
in special harmonic gauges. Instead in astronomy the positions of
stars and galaxies are determined from the data (luminosity, light
spectrum, angles) on the sky as living in a 4-dimensional
nearly-Galilei space-time with the celestial ICRS \cite{5} frame
considered as a "quasi-inertial frame" (all galactic dynamics is
Newtonian gravity), in accord with the assumed validity of the
cosmological and Copernican principles. Namely one assumes a
homogeneous and isotropic cosmological Friedmann-Robertson - Walker
solution of Einstein equations (the standard $\Lambda$CDM
cosmological model \cite{6}). In it the constant intrinsic
3-curvature of instantaneous 3-spaces is  nearly zero as implied by
the CMB data \cite{6}, so that Euclidean 3-spaces (and Newtonian
gravity) can be used. However, to reconcile all the data with this
4-dimensional reconstruction one must postulate the existence of
dark matter and dark energy as the dominant components of the
classical universe after the recombination 3-surface!

\bigskip

As a consequence, we need a metrology-oriented description of
non-inertial frames already in SR \cite{1,7}. This can be done with
the {\it 3+1 point of view} and the use of observer-dependent
Lorentz scalar radar 4-coordinates. Let us give the world-line
$x^{\mu}(\tau)$ of an arbitrary time-like observer carrying a
standard atomic clock: $\tau$ is an arbitrary monotonically
increasing function of the proper time of this clock. Then we give
an admissible 3+1 splitting of Minkowski space-time, namely a nice
foliation with space-like instantaneous 3-spaces $\Sigma_{\tau}$: it
is the mathematical idealization of a protocol for clock
synchronization (all the clocks in the points of $\Sigma_{\tau}$
sign the same time of the atomic clock of the observer). On each
3-space $\Sigma_{\tau}$ we choose curvilinear 3-coordinates
$\sigma^r$ having the observer as origin. These are the radar
4-coordinates $\sigma^A = (\tau; \sigma^r)$. If $x^{\mu} \mapsto
\sigma^A(x)$ is the coordinate transformation from the Cartesian
4-coordinates $x^{\mu}$ of a reference inertial observer to radar
coordinates, its inverse $\sigma^A \mapsto x^{\mu} = z^{\mu}(\tau
,\sigma^r)$ defines the {\it embedding} functions $z^{\mu}(\tau
,\sigma^r)$ describing the 3-spaces $\Sigma_{\tau}$ as embedded
3-manifold into Minkowski space-time. The induced 4-metric on
$\Sigma_{\tau}$ is the following functional of the embedding
${}^4g_{AB}(\tau ,\sigma^r) = [z^{\mu}_A\, \eta_{\mu\nu}\,
z^{\nu}_B](\tau ,\sigma^r)$, where $z^{\mu}_A = \partial\,
z^{\mu}/\partial\, \sigma^A$ and ${}^4\eta_{\mu\nu} = \sgn\, (+---)$
is the flat metric ($\sgn = \pm 1$ according to either the particle
physics $\sgn = 1$ or the general relativity $\sgn = - 1$
convention). While the 4-vectors $z^{\mu}_r(\tau ,\sigma^u)$ are
tangent to $\Sigma_{\tau}$, so that the unit normal $l^{\mu}(\tau
,\sigma^u)$ is proportional to $\epsilon^{\mu}{}_{\alpha
\beta\gamma}\, [z^{\alpha}_1\, z^{\beta}_2\, z^{\gamma}_3](\tau
,\sigma^u)$, we have $z^{\mu}_{\tau}(\tau ,\sigma^r) = [N\, l^{\mu}
+ N^r\, z^{\mu}_r](\tau ,\sigma^r)$ ($N(\tau ,\sigma^r) = \sgn\,
[z^{\mu}_{\tau}\, l_{\mu}](\tau ,\sigma^r)$ and $N_r(\tau ,\sigma^r)
= - \sgn\, g_{\tau r}(\tau ,\sigma^r)$ are the lapse and shift
functions).\medskip

The foliation is nice and admissible if it satisfies the conditions
\footnote{These conditions imply that global {\it rigid} rotations
are forbidden in relativistic theories. In Ref.\cite{1,8} there is
the expression of the admissible embedding corresponding to a 3+1
splitting of Minkowski space-time with parallel space-like
hyper-planes (not equally spaced due to a linear acceleration)
carrying differentially rotating 3-coordinates without the
coordinate singularity of the rotating disk. It is the first
consistent global non-inertial frame of this type.}: \hfill\break
 1) $N(\tau ,\sigma^r) > 0$ in every point of
$\Sigma_{\tau}$ (the 3-spaces never intersect, avoiding the
coordinate singularity of Fermi coordinates);\hfill\break
 2) $\sgn\, {}^4g_{\tau\tau}(\tau ,\sigma^r) > 0$, so to avoid the
 coordinate singularity of the rotating disk, and with the positive-definite 3-metric
${}^3g_{rs}(\tau ,\sigma^u) = - \sgn\, {}^4g_{rs}(\tau ,\sigma^u)$
having three positive eigenvalues (these are the M$\o$ller
conditions \cite{1,8});\hfill\break
 3) all the 3-spaces $\Sigma_{\tau}$ must tend to the same space-like
 hyper-plane at spatial infinity (so that there are always asymptotic inertial
observers to be identified with the fixed stars).\medskip

In the 3+1 point of view the 4-metric ${}^4g_{AB}(\tau ,\vec \sigma
)$ on $\Sigma_{\tau}$ has the components $\sgn\, {}^4g_{\tau\tau} =
N^2 - N_r\, N^r$, $- \sgn\, {}^4g_{\tau r} = N_r = {}^3g_{rs}\,
N^s$, ${}^3g_{rs} = - \sgn\, {}^4g_{rs} = \sum_{a=1}^3\,
{}^3e_{(a)r}\, {}^3e_{(a)s} = {\tilde \phi}^{2/3}\, \sum_{a=1}^3\,
e^{2\, \sum_{\bar b =1}^2\, \gamma_{\bar ba}\, R_{\bar b}}\,
V_{ra}(\theta^i)\, V_{sa}(\theta^i)$), where ${}^3e_{(a)r}(\tau
,\sigma^u)$ are cotriads on $\Sigma_{\tau}$, ${\tilde \phi}^2(\tau
,\sigma^r) = det\, {}^3g_{rs}(\tau ,\sigma^r)$ is the 3-volume
element on $\Sigma_{\tau}$, $\lambda_a(\tau ,\sigma^r) = [{\tilde
\phi}^{1/3}\, e^{\sum_{\bar b =1}^2\, \gamma_{\bar ba}\, R_{\bar
b}}](\tau ,\sigma^r)$ are the positive eigenvalues of the 3-metric
($\gamma_{\bar aa}$ are suitable numerical constants) and
$V(\theta^i(\tau ,\sigma^r))$ are diagonalizing rotation matrices
depending on three Euler angles. The components ${}^4g_{AB}$ or the
quantities $N$, $N_r$, $\gamma$, $R_{\bar a}$, $\theta^i$, play the
role of the {\it inertial potentials} generating the relativistic
apparent forces in the non-inertial frame. It can be shown
\cite{1,8} that the Newtonian inertial potentials are hidden in the
functions $N$, $N_r$ and $\theta^i$. The extrinsic curvature
${}^3K_{rs}(\tau, \sigma^u) = [{1\over {2\, N}}\, (N_{r|s} + N_{s|r}
- \partial_{\tau}\, {}^3g_{rs})](\tau, \sigma^u)$, describing the
{\it shape} of the instantaneous 3-spaces of the non-inertial frame
as embedded 3-manifolds of Minkowski space-time, is a secondary
inertial potential functional of the independent inertial potentials
${}^4g_{AB}$.\medskip

Instead in GR the 4-metric is described by ten dynamical fields
${}^4g_{\mu\nu}(x)$: it is not only a (pre)potential for the
gravitational field but also determines the chrono-geometrical
structure of space-time through the line element $ds^2 =
{}^4g_{\mu\nu}\, dx^{\mu}\, dx^{\nu}$ (it teaches relativistic
causality to the other fields) \footnote{The ACES mission of ESA
\cite{9} will give the first precision measurement of the
gravitational redshift of the geoid, namely of the $1/c^2$
deformation of Minkowski light-cone caused by the geo-potential. In
every quantum field theory, where the definition of the Fock space
requires the use of the fixed light-cone of the background, this is
a non-perturbative effect requiring the resummation of the
perturbative expansion.}. To get its Hamiltonian description in the
quoted restricted class of space-times  the same 3+1 point of view
and the radar 4-coordinates employed in SR can be used: this allows
to separate the inertial (gauge) degrees of freedom of the
gravitational field (playing the role of inertial potentials) from
the dynamical tidal ones. But now the admissible embeddings $x^{\mu}
= z^{\mu}(\tau, \sigma^r)$ are not dynamical variables: instead
their gradients $z^{\mu}_A(\tau, \sigma^r)$ give the transition
coefficient from radar to world 4-coordinates, ${}^4g_{AB}(\tau,
\sigma^r) = [z^{\mu}_A\, z^{\nu}_B](\tau, \sigma^r)\,
{}^4g_{\mu\nu}(z(\tau, \sigma^r))$. As shown in Ref.\cite{10}, the
dynamical nature of space-time implies that each solution of
Einstein's equations dynamically selects a preferred 3+1 splitting
of the space-time, namely in GR the instantaneous 3-spaces (and
therefore the associated clock synchronization convention) are
dynamically determined. Now the extrinsic curvature of the 3-spaces
will be a mixture of dynamical (tidal) pieces and inertial gauge
variables playing the role of inertial potentials.

\bigskip

The description of isolated systems (particles, strings, fields,
fluids) admitting a Lagrangian formulation in the non-inertial
frames of SR is done by means of {\it parametrized Minkowski
theories} \cite{1,7}. The matter variables are replaced with new
ones knowing the 3-spaces $\Sigma_{\tau}$. For instance a
Klein-Gordon field $\tilde \phi (x)$ will be replaced with
$\phi(\tau ,\sigma^r) = \tilde \phi (z(\tau ,\sigma^r))$; the same
for every other field. Instead for a relativistic particle with
world-line $x^{\mu}(\tau )$ we must make a choice of its energy
sign: then it will be described by 3-coordinates $\eta^r(\tau )$
defined by the intersection of the world-line with $\Sigma_{\tau}$:
$x^{\mu}(\tau ) = z^{\mu}(\tau ,\eta^r(\tau ))$. Differently from
all the previous approaches to relativistic mechanics, the dynamical
configuration variables are the 3-coordinates $\eta^r_i(\tau)$ and
not the world-lines $x^{\mu}_i(\tau)$ (to rebuild them in an
arbitrary frame we need the embedding defining that frame!). Then
the matter Lagrangian is coupled to an external gravitational field
and the external 4-metric is replaced with the 4-metric $g_{AB}(\tau
,\sigma^r)$ of an admissible 3+1 splitting of Minkowski space-time.
With this procedure we get a Lagrangian depending on the given
matter and on the embedding $z^{\mu}(\tau ,\sigma^r)$, which is
invariant under {\it frame-preserving diffeomorphisms}. As a
consequence, there are four first-class constraints (an analogue of
the super-Hamiltonian and super-momentum constraints of canonical
gravity) implying that the embeddings $z^{\mu}(\tau ,\sigma^r)$ are
{\it gauge variables}, so that all the admissible non-inertial or
inertial frames are gauge equivalent, namely physics does {\it not}
depend on the clock synchronization convention and on the choice of
the 3-coordinates $\sigma^r$: only the appearances of phenomena
change by changing the notion of instantaneous 3-space. Even if the
gauge group is formed by the frame-preserving diffeomorphisms, the
matter energy-momentum tensor allows the determination of the ten
conserved Poincare' generators $P^{\mu}$ and $J^{\mu\nu}$ (assumed
finite) of every configuration of the system.

 \medskip

If we restrict ourselves to inertial frames, we can define the {\it
inertial rest-frame instant form of dynamics for isolated systems}
by choosing the 3+1 splitting corresponding to the intrinsic
inertial rest frame of the isolated system centered on an inertial
observer: the instantaneous 3-spaces, named Wigner 3-space due to
the fact that the 3-vectors inside them are Wigner spin-1 3-vectors
\cite{7}, are orthogonal to the conserved 4-momentum $P^{\mu}$ of
the configuration. In Ref.\cite{1} there is the extension to
admissible {\it non-inertial rest frames}, where $P^{\mu}$ is
orthogonal to the asymptotic space-like hyper-planes to which the
instantaneous 3-spaces tend at spatial infinity. This non-inertial
family of 3+1 splittings is the only one admitted by the
asymptotically Minkowskian space-times covered by canonical gravity
formulation discussed below. \medskip

The framework of the inertial rest frame allowed the solution of the
following old open problems:\medskip

A) The explicit form of the Lorentz boosts for some interacting
systems \cite{11,12}.\medskip

B) The classification of the relativistic collective variables,
replacing the Newtonian center of mass, that can be built in terms
of the Poincare' generators and their non measurability due to the
non-local character of such generators (they know the whole 3-space
$\Sigma_{\tau}$) \cite{1,13}.\medskip

C) The description of every isolated system as a decoupled
(non-measurable) canonical non-covariant (Newton-Wigner) center of
mass carrying a pole-dipole structure: the invariant mass and the
rest spin of the system expressed in terms of Wigner-covariant
relative variables inside the Wigner 3-spaces
\cite{1,13,14,15}.\medskip

D) The formulation of classical relativistic atomic physics
\cite{11,14} (the electro-magnetic field in the radiation gauge plus
charged scalar particles with Grassmann-valued electric charges to
regularize the self-energies) and the identification of the Darwin
potential at the classical level by evading Haag's theorem.\medskip

E) A new formulation of {\it relativistic quantum mechanics}
\cite{15} englobing all the known results about relativistic bound
states and a first formulation of {\it relativistic entanglement}
taking into account the {\it non-locality and spatial
non-separability coming from the Poincare' group}.

\bigskip

Let us now consider Einstein's general relativity where space-time
is dynamical. Since all the properties of the standard model of
elementary particles are connected with properties of the
representations of the Poincare' group in inertial frames of
Minkowski space-time, we shall restrict ourselves to globally
hyperbolic, asymptotically Minkowskian at spatial infinity,
topologically trivial space-times, for which a well defined
Hamiltonian formulation of gravity is possible if we replace the
Hilbert action with the ADM one. The 4-metric tends in a suitable
way to the flat Minkowski 4-metric ${}^4\eta_{\mu\nu}$ at spatial
infinity: having an {\it asymptotic} Minkowskian background we can
avoid to split the 4-metric in the bulk in a background plus
perturbations in the weak field limit. In these space-times we can
use admissible 3+1 splittings and observer-dependent radar
4-coordinates. Since tetrad gravity is more natural for the coupling
of gravity to the fermions, the 4-metric is decomposed in terms of
cotetrads, ${}^4g_{AB} = E_A^{(\alpha)}\,
{}^4\eta_{(\alpha)(\beta)}\, E^{(\beta)}_B$ \footnote{$(\alpha)$ are
flat indices; the cotetrads $E^{(\alpha)}_A$ are the inverse of the
tetrads $E^A_{(\alpha)}$ connected to the world tetrads by
$E^{\mu}_{(\alpha)}(x) = z^{\mu}_A(\tau, \sigma^r)\,
E^A_{(\alpha)}(z(\tau, \sigma^r))$.}, and the ADM action, now a
functional of the 16 fields $E^{(\alpha)}_A(\tau, \sigma^r)$, is
taken as the action for ADM tetrad gravity. In tetrad gravity the
diffeonorphism group is enlarged with the O(3,1) gauge group of the
Newman-Penrose approach (the extra gauge freedom acting on the
tetrads in the tangent space of each point of space-time and
reducing from 16 to 10 the number of independent fields like in
metric gravity). This leads to an interpretation of gravity based on
a congruence of time-like observers endowed with orthonormal
tetrads: in each point of space-time the time-like axis is the  unit
4-velocity of the observer, while the spatial axes are a (gauge)
convention for observer's gyroscopes. This framework was developed
in the works in Refs.\cite{16}.

\medskip

In these space-times we assume direction-independent boundary
conditions on the 4-metric and its conjugate momenta able to kill
{\it super-translations} \cite{17}, so that the SPI group of
asymptotic symmetries \cite{18} is reduced to the ADM Poincare'
group with the generators $P^A_{ADM}$, $J^{AB}_{ADM}$ given as
boundary conditions. It turns out that the admissible 3+1 splittings
are the {\it non-inertial rest frames} (with the 3-spaces
asymptotically orthogonal to the ADM 4-momentum; $P^r_{ADM} \approx
0$ is the rest-frame condition) of the 3-universe with a mass and a
rest spin fixed by the boundary conditions \footnote{Therefore there
are asymptotic inertial observers to be identified with the fixed
stars of star catalogues. If $\epsilon^{\mu}_A$ are a set of
asymptotic flat tetrads, the simplest embedding adapted to the 3+1
splitting of space-time is $x^{\mu} = z^{\mu}(\tau, \sigma^r) =
x^{\mu}(\tau) + \epsilon^{\mu}_r\, \sigma^r = x^{\mu}_o +
\epsilon^{\mu}_A\, \sigma^A$ and we have ${}^4g_{AB}(\tau, \sigma^r)
= \epsilon^{\mu}_A\, \epsilon^{\nu}_B\, {}^4g_{\mu\nu}(x)$.}. In
absence of matter Christodoulou - Klainermann space-times \cite{19}
are compatible with this description. With this kind of formalism we
can get a deparametrization of general relativity: if we switch off
the Newton constant and we choose the flat Minkowski 4-metric in
Cartesian coordinates as solution of Einstein's equations, we get
the description of the matter present in the 3-universe in the
non-inertial rest frames of Minkowski space-time with the weak ADM
Poincare' group collapsing in the Poincare' group of particle
physics. As shown in Refs.\cite{16}, with the previous boundary
conditions the DeWitt surface term at spatial infinity in the Dirac
Hamiltonian turns out to be the {\it strong} ADM energy (a flux
through a 2-surface at spatial infinity), which is equal to the {\it
weak} ADM energy (expressed as a volume integral over the 3-space)
plus constraints. Therefore in this family of space-times there is
{\it not a frozen picture}, like in the family of spatially compact
without boundary space-times considered in loop quantum gravity,
where the Dirac Hamiltonian is a combination of constraints.

\medskip

In this framework the configuration variables are cotetrads, which
are connected to cotetrads adapted to the 3+1 splitting of
space-time (so that the adapted time-like tetrad is the unit normal
to the 3-space $\Sigma_{\tau}$) by standard Wigner boosts for
time-like vectors of parameters $\varphi_{(a)}(\tau, \sigma^r)$,
$a=1,2,3$: $E_A^{\alpha)} = L^{(\alpha)}{}_{(\beta)}(
\varphi_{(a)})\, {\buildrel o\over E}_A^{(\beta)}$. The adapted
cotetrads have the following expression in terms of cotriads
${}^3e_{(a)r}$ on $\Sigma_{\tau}$ and of  the lapse $N = 1 + n$ and
shift $n_{(a)} = N^r\, {}^3e_{(a)r}$ functions: ${\buildrel o\over
E}_{\tau}^{(o)} = 1 + n$, ${\buildrel o\over E}_r^{(o)} = 0$,
${\buildrel o\over E}_{\tau}^{(a)} = n_{(a)}$, ${\buildrel o\over
E}_r^{(a)} = {}^3e_{(a)r}$. The 4-metric becomes ${}^4g_{\tau\tau} =
\sgn\, [(1 + n)^2 - \sum_a\, n^2_{(a)}]$, ${}^4g_{\tau r} = - \sgn\,
\sum_a\, n_{(a)}\, {}^3e_{(a)r}$, ${}^4g_{rs} = - \sgn\, {}^3g_{rs}
= - \sgn\, \sum_a\, {}^3e_{(a)r}\, {}^3e_{(a)s}$. The 16
configurational variables in the ADM action are $\varphi_{(a)}$, $1
+ n$, $n_{(a)}$, ${}^3e_{(a)r}$. There are ten primary constraints
(the vanishing of the 7 momenta of boosts, lapse and shift variables
plus three constraints describing the rotation on the flat indices
$(a)$ of the cotriads) and four secondary ones (the
super-Hamiltonian and super-momentum constraints): all of them are
first class in the phase space spanned by 16+16 fields. This implies
that there are 14 gauge variables describing {\it inertial effects}
and 2 canonical pairs of physical degrees of freedom describing the
{\it tidal effects} of the gravitational field (namely gravitational
waves in the weak field limit). In this canonical basis only the
momenta ${}^3\pi^r_{(a)}$ conjugated to the cotriads are not
vanishing.
\medskip

Then in Ref.\cite{18} we have found a canonical transformation to a
canonical basis adapted to ten of the first class constraints. It
implements the York map of Ref.\cite{20} and diagonalizes the
York-Lichnerowicz approach \cite{21}. Its final form is
($\alpha_{(a)}(\tau, \sigma^r)$ are angles)

\bea
 &&\begin{minipage}[t]{4 cm}
\begin{tabular}{|ll|ll|l|l|l|} \hline
$\varphi_{(a)}$ & $\alpha_{(a)}$ & $n$ & ${\bar n}_{(a)}$ &
$\theta^r$ & $\tilde \phi$ & $R_{\bar a}$\\ \hline
$\pi_{\varphi_{(a)}} \approx0$ &
 $\pi^{(\alpha)}_{(a)} \approx 0$ & $\pi_n \approx 0$ & $\pi_{{\bar n}_{(a)}} \approx 0$
& $\pi^{(\theta )}_r$ & $\pi_{\tilde \phi} = {{c^3}\over {12\pi G}}\, {}^3K$ & $\Pi_{\bar a}$ \\
\hline
\end{tabular}
\end{minipage}\nonumber \\
 &&{}\nonumber \\
 &&{}\nonumber \\
 &&{}^3e_{(a)r} = R_{(a)(b)}(\alpha_{(c)})\, {}^3{\bar e}_{(b)r} =
 R_{(a)(b)}(\alpha_{(c)})\, V_{rb}(\theta^i)\,
 {\tilde \phi}^{1/3}\, e^{\sum_{\bar a}^{1,2}\, \gamma_{\bar aa}\, R_{\bar a}},\nonumber \\
 &&{}^4g_{\tau\tau} = \sgn\, [(1 + n)^2 - \sum_a\, {\bar n}^2_{(a)}],
 \qquad {}^4g_{\tau r} = - \sgn\, {\bar
 n}_{(a)}\, {}^3{\bar e}_{(a)r},\nonumber \\
 &&{}^4g_{rs} = - \sgn\, {}^3g_{rs} = - \sgn\, {\tilde \phi}^{2/3}\,
 \sum_a\, V_{ra}(\theta^i)\, V_{sa}(\theta^i)\,
 e^{2\, \sum_{\bar a}^{1,2}\, \gamma_{\bar aa}\, R_{\bar
 a}},\nonumber \\
 &&{}\nonumber \\
 \eea

In this York canonical basis the {\it inertial effects} are
described by the arbitrary gauge variables $\alpha_{(a)}$,
$\varphi_{(a)}$, $1 + n$, ${\bar n}_{(a)}$, $\theta^i$, ${}^3K$,
while the {\it tidal effects}, i.e. the physical degrees of freedom
of the gravitational field, by the two canonical pairs $R_{\bar a}$,
$\Pi_{\bar a}$, $\bar a =1,2$. The momenta $\pi_r^{(\theta)}$ and
the 3-volume element $\tilde \phi = \sqrt{det\, {}^3g_{rs}}$ have to
be found as solutions of the super-momentum and super-hamiltonian
(i.e. the Lichmerowicz equation) constraints, respectively. The
gauge variables $\alpha_{(a)}$, $\varphi_{(a)}$ parametrize the
extra O(3,1) gauge freedom of the tetrads (the gauge freedom for
each observer to choose three gyroscopes as spatial axes and to
choose the law for their transport along the world-line). The gauge
angles $\theta^i$ (i.e. the director cosines of the tangents to the
three coordinate lines in each point of $\Sigma_{\tau}$) describe
the freedom in the choice of the 3-coordinates $\sigma^r$ on each
3-space: their fixation implies the determination of the shift gauge
variables ${\bar n}_{(a)}$, namely the appearances of
gravito-magnetism in the chosen 3-coordinate system.\medskip

The final basic gauge variable is a momentum, namely the trace
${}^3K(\tau ,\sigma^r)$ of the extrinsic curvature (also named the
{\it York time}) of the non-Euclidean 3-space $\Sigma_{\tau}$. The
Lorentz signature of space-time implies that ${}^3K$ is a momentum
variable: it is a time coordinate, while $\theta^i$ are spatial
coordinates. Differently from SR ${}^3K$ is {\it an independent
inertial gauge variable describing the remnant in GR of the freedom
in clock synchronization}! The other components of the extrinsic
curvature are dynamically determined. This gauge variable has no
Newtonian counterpart (the Euclidean 3-space is absolute), because
its fixation determines the final shape of the non-Euclidean
3-space. Moreover this gauge variable gives rise to a negative
kinetic term in the weak ADM energy ${\hat E}_{ADM}$, vanishing only
in the gauges ${}^3K(\tau, \vec \sigma) = 0$ \cite{18}.

\medskip

In the York canonical basis the Hamilton equations generated by the
Dirac Hamiltonian $H_D = {\hat E}_{ADM} + (constraints)$ are divided
in four groups: A) the contracted Bianchi identities, namely the
evolution equations for $\tilde \phi$ and $\pi_i^{(\theta)}$ (they
say that given a solution of the constraints on a Cauchy surface, it
remains a solution also at later times); B) the evolution equation
for the four basic gauge variables $\theta^i$ and ${}^3K$: these
equations determine the lapse and the shift functions once the basic
gauge variables are fixed; C) the evolution equations for the tidal
variables $R_{\bar a}$, $\Pi_{\bar a}$; D) the Hamilton equations
for matter, when present. Once a gauge is completely fixed, the
Hamilton equations become deterministic. Given a solution of the
super-momentum and super-Hamiltonian constraints and the Cauchy data
for the tidal variables on an initial 3-space, we can find a
solution of Einstein's equations in radar 4-coordinates adapted to a
time-like observer.

\medskip

In the first paper of Ref.\cite{22}, we studied the coupling of N
charged scalar particles plus the electro-magnetic field to ADM
tetrad gravity  in this class of asymptotically Minkowskian
space-times without super-translations. To regularize the
self-energies both the electric charge and the sign of the energy of
the particles are Grassmann-valued. The introduction of the
non-covariant radiation gauge allows to reformulate the theory in
terms of transverse electro-magnetic fields and to extract the
generalization of the action-at-a- distance Coulomb interaction
among the particles in the Riemannian instantaneous 3-spaces of
global non-inertial frames. After the reformulation of the whole
system in the York canonical basis, we give the restriction of the
Hamilton equations and of the constraints to the family of {\it
non-harmonic 3-orthogonal} Schwinger time gauges, in which the
instantaneous Riemannian 3-spaces have a non-fixed trace ${}^3K$ of
the extrinsic curvature but a diagonal 3-metric. This family of
gauges is determined by the gauge fixings $\theta^i(\tau, \sigma^r)
\approx 0$ and ${}^3K(\tau, \sigma^r) \approx (arbitrary\,
numerical\, function)$.

\medskip

In the second paper of Ref.\cite{22} it was shown that in this
family of non-harmonic 3-orthogonal Schwinger gauges it is possible
to define a consistent {\it linearization} of ADM canonical tetrad
gravity plus matter in the weak field approximation, to obtain a
formulation of Hamiltonian Post-Minkowskian gravity with non-flat
Riemannian 3-spaces and asymptotic Minkowski background. This means
that the 4-metric tends to the asymptotic Minkowski metric at
spatial infinity, ${}^4g_{AB}\, \rightarrow {}^4\eta_{AB}$. The
decomposition ${}^4g_{AB} = {}^4\eta_{AB} + {}^4h_{(1)AB}$, with a
first order perturbation ${}^4h_{(1)AB}$ vanishing at spatial
infinity, is only used for calculations, but has no intrinsic
meaning. Moreover, due to the presence of a ultra-violet cutoff for
matter, we can avoid to make Post-Newtonian expansions, namely we
get fully relativistic expressions. We have found solutions for the
first order quantities $\pi^{(\theta)}_{(1)r}$, $\tilde \phi = 1 +
6\, \phi_{(1)}$, $1 + n_{(1)}$, ${\bar n}_{(1)(a)}$ (the
action-at-a-distance part of the gravitational interaction). Then we
can show that the tidal variables $R_{\bar a}$ satisfy a wave
equation $\Box\, R_{\bar a} = (known\, functional\, of\, matter)$
with the D'Alambertian associated to the asymptotic Minkowski
4-metric. Therefore, by using a no-incoming radiation condition
based on the asymptotic Minkowski light-cone, we get a description
of gravitational waves in these non-harmonic gauges, which can be
connected to generalized TT(transverse traceless) gauges, as
retarded functions of the matter. These gravitational waves do not
propagate in inertial frames of the background (like it happens in
the standard harmonic gauge description), but in non-Euclidean
instantaneous 3-spaces differing from Euclidean 3-spaces at the
first order (their intrinsic 3-curvature is determined by the
gravitational waves) and dynamically determined by the linearized
solution of Einstein equations. These 3-spaces have a first order
extrinsic curvature (with ${}^3K_{(1)}(\tau, \sigma^r)$ describing
the clock synchronization convention) and a first order modification
of Minkowski light-cone.
\medskip

We can write explicitly the linearized Hamilton equations for the
particles and for the electro-magnetic field: among the forces there
are both the inertial potentials and the gravitational waves. In the
third paper of Ref.\cite{22} we disregarded electro-magnetism and we
studied the non-relativistic limit of the particle equations. We
found  that the particle 3-coordinates $\eta^r_i(\tau = ct) =
{\tilde \eta}_i^r(t)$ satisfy the equation  $m\, {{d^2 {\tilde
\eta}^r_i(t)}\over {dt^2}} = \sum_{j \not= i}\, F^r_{Newton}({\vec
{\tilde \eta}}_i(t) - {\vec {\tilde \eta}}_j(t)) + {1\over c}\, {{d
{\tilde \eta}^r_i(t)}\over {dt}}\, \Big({1\over {\triangle}}\, c^2\,
\partial^2_{\tau}\, {}^3K_{(1)}(\tau = ct, \vec \sigma)\Big)
{|}_{\vec \sigma = {\vec {\tilde \eta}}_i(t)}$, where ${\vec
F}_{Newton}$ is the Newton gravitational force. Therefore the
(arbitrary in these gauges) double rate of change in time of the
trace of the extrinsic curvature creates a post-Newtonian damping
(or anti-damping since the sign of ${}^3K_{(1)}$ is not fixed)
effect on the motion of particles. This is a inertial effect (hidden
in the lapse function) not existing in Newton theory where the
Euclidean 3-space is absolute. In the 2-body case we get that for
Keplerian circular orbits of radius $r$ the modulus of the relative
3-velocity can be written in the form $\sqrt{{{G\, (m + \triangle\,
m(r))}\over r}}$ with $\triangle\, m(r)$ function only of
${}^3K_{(1)}$. Now the rotation curves of galaxies (see
Ref.\cite{23} for a review) imply that this quantity goes to
constant for large $r$ (instead of vanishing): as a consequence
$\triangle\, m(r)$ is interpreted as a {\it dark matter halo} around
the galaxy. With our approach this dark matter would be a {\it
relativistic inertial effect} consequence of the non-Euclidean
nature of 3-space. This option would differ:  1) from the
non-relativistic MOND approach \cite{24} (where one modifies Newton
equations); 2) from modified gravity theories like the $f(R)$ ones
(see for instance Refs.\cite{25}; here one gets a modification of
the Newton potential); 3) from postulating the existence of WIMP
particles \cite{26}.

\medskip

Since, as already said, at the experimental level {\it the
description of baryon matter is intrinsically coordinate-dependent},
namely is connected with the conventions used by physicists,
engineers and astronomers for the modeling  of space-time. As a
consequence of the dependence on coordinates of the description of
matter, our proposal for solving the gauge problem in our
Hamiltonian framework with non-Euclidean 3-spaces is to choose a
gauge (i.e. a 4-coordinate system) in non-modified Einstein gravity
which is in agreement with the observational conventions in
astronomy. Since ICRS \cite{5} has diagonal 3-metric, our
3-orthogonal gauges are a good choice. We are left with the inertial
gauge variable ${}^3{\cal K}_{(1)} = {1\over {\triangle}}\,
{}^3K_{(1)}$ not existing in Newtonian gravity. As already said the
suggestion is to try to fix ${}^3{\cal K}_{(1)}$ in such a way to
eliminate dark matter as much as possible, by reinterpreting it as a
relativistic inertial effect induced by the shift from Euclidean
3-spaces to non-Euclidean ones (independently from cosmological
assumptions). As a consequence, ICRS should be reformulated not as a
{\it quasi-inertial} reference frame in Galilei space-time, but as a
reference frame in a PM space-time with ${}^3K$ (i.e. the clock
synchronization convention) deduced from the data connected to dark
matter. Then automatically BCRS would be its quasi-Minkowskian
approximation  for the Solar System. This point of view could also
be useful for the ESA GAIA mission (cartography of the Milky Way)
\cite{27}.
\bigskip

In conclusion the Hamiltonian formulation of Einstein theory done in
a form which takes into account the problem of 3-space (i.e. of
clock synchronization) and the coordinate-dependent description of
matter (i.e. metrology) opens a new scenario for dark matter.
Besides looking for other experimental signatures of the York time,
we also have the possibility to explore its role in the
back-reaction approach \cite{28} to dark energy, according to which
dark energy is a byproduct of the non-linearities of general
relativity when one considers spatial mean values on large scales to
get a cosmological description of the universe taking into account
the inhomogeneity of the observed universe. In the York canonical
basis all the relevant quantities are 3-scalars and  it is possible
to study the mean value of nearly all the Hamilton equations.

\end{document}